\documentclass{pasa}%

\usepackage{hyperref}
\usepackage{natbib}
\usepackage{graphicx}

\usepackage{pdflscape}
\usepackage{afterpage}

\usepackage{calrsfs}

\usepackage{soul}

\title[Enabling remote search for fast transient events]{Enabling near real-time remote search for fast transient events with lossy data compression}

\author[Vohl et al.]{Dany Vohl$^{1,2}$\thanks{Corresponding author: dvohl@swin.edu.au}, Tyler Pritchard$^{1,3}$, Igor Andreoni$^{1,3,4,5}$, Jeffrey Cooke$^{1,3,5}$, and Bernard Meade$^{1,6}$
\affil{$^1$Centre for Astrophysics and Supercomputing, Swinburne University of Technology, Hawthorn 3122, Australia \\
$^{2}$Advanced Visualisation Laboratory, Digital Research \& Innovation Capability Platform, Swinburne University of Technology, Hawthorn, Australia \\
$^{3}$Australian Research Council Centre of Excellence for All-sky Astrophysics (CAASTRO), Australia \\
$^{4}$Australian Astronomical Observatory, North Ryde 2113, Australia \\
$^{5}$Australian Research Council Centre of Excellence for Gravitational Wave Discovery (OzGrav), Australia \\
$^{6}$The University of Melbourne, Parkville 3010, Australia}%
}%

\jid{PASA}
\doi{10.1017/pas.\the\year.xxx}
\jyear{\the\year}

\usepackage{aas_macros}
\usepackage{hyperref} 
\hypersetup{colorlinks,citecolor=blue,linkcolor=blue,urlcolor=blue}

\begin{document}

\begin{frontmatter}
\maketitle
\begin{abstract}
We present a systematic evaluation of JPEG2000 (ISO/IEC 15444) as a transport data format to enable rapid remote searches for fast transient events as part of the Deeper Wider Faster program (DWF).  DWF uses $\sim$20 telescopes from radio to gamma-rays to perform simultaneous and rapid-response follow-up searches for fast transient events on millisecond-to-hours timescales.  DWF search demands have a set of constraints that is becoming common amongst large collaborations.  Here, we focus on the rapid optical data component of DWF led by the Dark Energy Camera (DECam) at Cerro Tololo Inter-American Observatory (CTIO).  Each DECam image has 70 total coupled-charged devices saved as a $\sim$1.2 gigabyte FITS file.  Near real-time data processing and fast transient candidate identifications -- in minutes for rapid follow-up triggers on other telescopes -- requires computational power exceeding what is currently available on-site at CTIO.  In this context, data files need to be transmitted rapidly to a foreign location for supercomputing post-processing, source finding, visualization and analysis.  This step in the search process poses a major bottleneck, and reducing the data size helps accommodate faster data transmission.  To maximise our gain in transfer time and still achieve our science goals, we opt for lossy data compression --- keeping in mind that raw data is archived and can be evaluated at a later time.  We evaluate how lossy JPEG2000 compression affects the process of finding transients, and find only a negligible effect for compression ratios up to $\sim$25:1.  We also find a linear relation between compression ratio and the mean estimated data transmission speed-up factor.  Adding highly customized compression and decompression steps to the science pipeline considerably reduces the transmission time --- validating its introduction to the DWF science pipeline and enabling science that was otherwise too difficult with current technology.  

\end{abstract}
\end{frontmatter}

\section{INTRODUCTION }
\label{sec:introduction}
Data compression, issued from the field of information theory \citep{shannon}, is the practice of transforming a data file into a more compact representation of itself.  Data compression increases the amount of data that can be stored on disk (or other storage medium), and helps reduce the time required to transmit data over a noisy network.  It has been used to minimize the volume of astronomical data since the 1970s, and has continued to be developed and used ever since \cite[e.g.][]{labrum1975radioheliography, white-1994, Pence2000, Pence2011, Tomasi2016A&C....16...88T}.  Two main categories of compression exist: lossless and lossy compression.  Lossless compression yields smaller compression ratios than lossy compression, but permits one to retrieve the exact original data after decompression.  Lossy compression results in an approximation of the original data, requiring one to assess the decompressed data, but can still enable sound scientific analysis.  

In recent years, ``Big Data'' issues have become more prominent for large astronomical projects.  The main characteristics of ``Big Data'' are often described as \emph{volume}, \emph{velocity}, and \emph{variety} \citep{Wu2014BDR}.  The volume refers to the amount of information that systems must ingest, process and disseminate.  The velocity refers to the speed at which information grows or disappears.  Finally, the variety refers to the diversity of data sources and formats.  While the variety of formats is generally represented by a limited set of options for a given sub-field [e.g. FITS \citep[][]{Wells1981A&AS...44..363W}, HDF5 \citep[][]{Folk:2011:OHT:1966895.1966900}], the volume and velocity have a direct impact in modern astronomy. 

Recently, a large collaboration of astronomers has been taking part in the Deeper Wider Faster (DWF) initiative (Cooke et al., in prep.) --- a remote and time-critical observation program. DWF is a coordinated multi-wavelength observing effort, that includes $\gtrsim20$ facilities located worldwide and in space, which aims to identify, in near real-time, fast transient events on millisecond-to-hours timescales.  Such events include Fast Radio Bursts \citep[FRBs,][]{Lorimer2007Sci...318..777L}, Gamma-ray bursts \citep[GRBs,][]{Klebesadel1973ApJ...182L..85K}, kilonov\ae \ \citep{Li1998ApJ...507L..59L} and ultra-luminous X-Ray sources \citep{Miller2004ApJ...614L.117M}. 

To cover a wide range of wavelengths, DWF uses a variety of instruments including the Dark Energy Camera \citep[DECam;][]{Diehl2012PhPro..37.1332D, Flaugher2012SPIE.8446E..11F, Flaugher2015AJ....150..150F} installed at the Cerro Tololo Inter-American Observatory (CTIO), the Molonglo Observatory Synthesis Telescope (MOST), the NASA SWIFT Space Telescope, the Parkes observatory, the Antarctica Schmidt telescopes (AST3), the Gemini Observatory, Southern African Large Telescope (SALT), the Anglo-Australian Telescope (AAT), the SkyMapper telescope, the Zadko Telescope, the Rapid Eye Mount telescope (REM), and the Laser Interferometer Gravitational-Wave Observatories (LIGO).  In the present paper, we focus our attention on DECam and its data products. 

\begin{figure}[!ht]
\centering
\includegraphics[width=8cm]{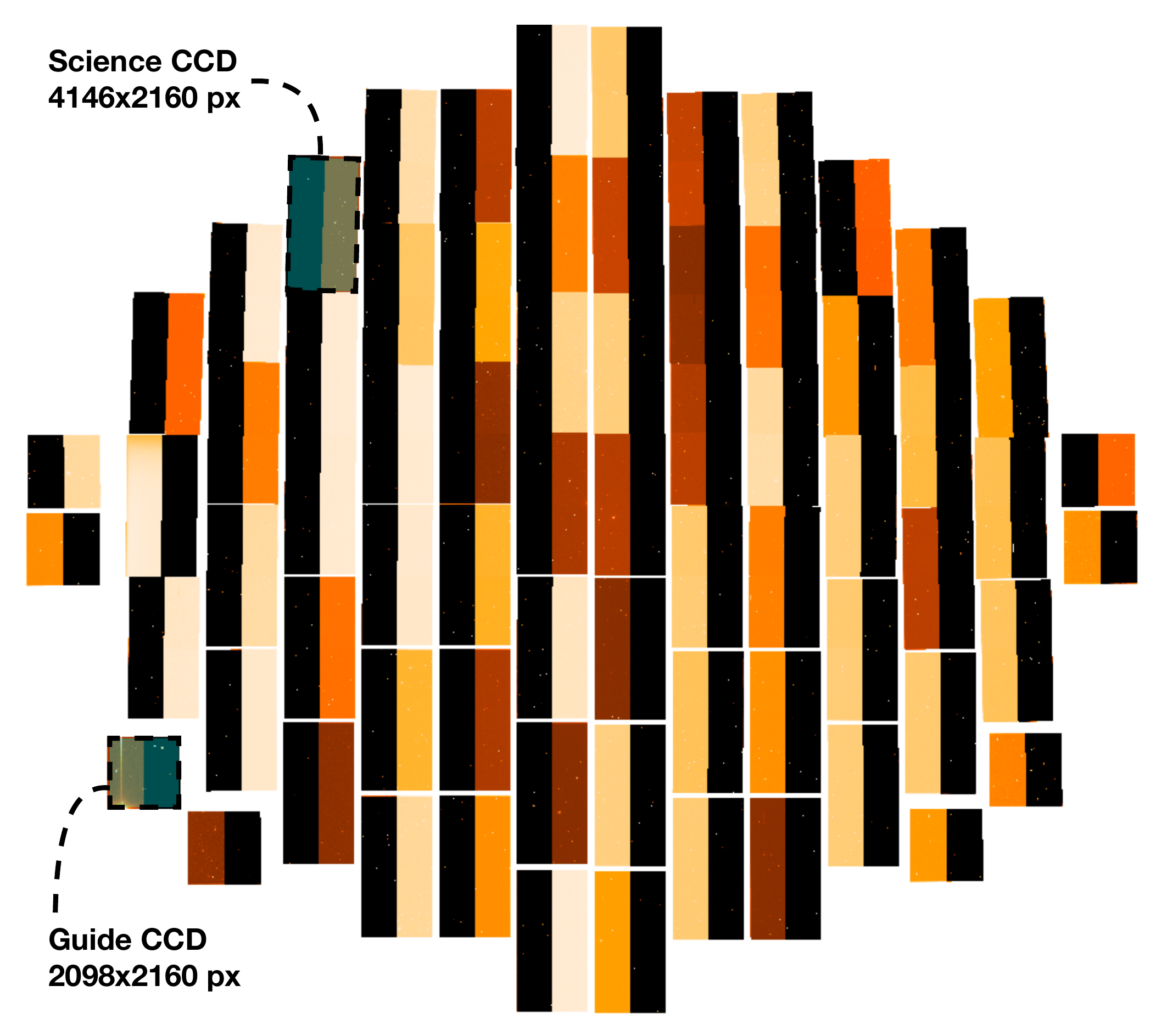}
\caption{Example of a raw and uncalibrated mosaic image, as captured by the 62 science CCDs and 8 guides CCDs of DECam.  Each science CCD is of dimension $4146\times2160$ pixels, while each guide CCD contains $2098\times2160$ pixels.  Each pixel is encoded as a 32-bit integer, resulting to $\sim1.2$ GB of storage space for the whole mosaic. Each CCD has two amplifiers, providing the ability to read the pixel arrays using either or both amplifiers. Each CCD of the uncalibrated image displays a darker and a lighter side, corresponding to the regions covered by each amplifier. The mosaic was visualised with SAOImage DS9 \citep{ds9-2000ascl.soft03002S} using the heat colour map. The blue masks and dashed lines highlight the size of a science and guide CCD respectively.}
\label{fig::decam}
\end{figure}

Data generated with DECam are of imposing size.  DECam is composed of a mosaic of 70 coupled-charged devices (CCDs; Figure \ref{fig::decam}), including 62 science CCDs and 8 guide CCDs. Each science CCD is of dimension $4146\times2160$ pixels, while each guide CCD contains $2098\times2160$ pixels.  A mosaic image is saved as a FITS file, where each pixel of an image is stored as a 32-bit integer ({\tt BITPIX}).  This results is a data file requiring $\sim1.2$ gigabyte (GB) of storage space (pre-processing).  

During a DWF observation campaign (hereafter {\em run}), data files are acquired every 40 seconds from a continuous stream of 20-second exposures, each followed by a 20-second readout time provided by the DECam electronics.  This observing cadence, and the high sensitivity of DECam, enables the DWF campaign to search for fast transients, while maintaining survey depth and time on sky.  Each field is observed simultaneously for 1 to 2 hours per night by several observatories, as a result of field constraints imposed by the coincident visibility of DECam in Chile and Parkes and Molonglo in Australia.  As a result, around 100 to 200 DECam optical images are acquired per field per night during a run (and three to seven fields per night).

To search for transient candidates in near real-time requires computational power that exceeds what is currently available on-site at CTIO.  In this context, data files constantly need to be transmitted to a suitable location for post-processing, source finding, visualization and analysis \citep{Meade2017arXiv170401281M, Andreoni-accepted}. The Green II supercomputer\footnote{http://supercomputing.swin.edu.au} at Swinburne University of Technology in Australia provides the computational power necessary for the main DWF goals.  However, transmission of large amount of raw data from CTIO to Australia, where our group is located, represents a major bottleneck.  To accelerate this process, we integrate data compression as part of the science pipeline.  To maximise our gain in transmission time, we choose to use lossy compression --- keeping in mind that raw data is archived and can be evaluated at a later time. 

\subsection{JPEG2000 and lossy data compression}
\label{section::jpeg2000}

Several lossy compression techniques have been proposed for astronomical images over the years. These include compression techniques based on Rice compression \citep{Pence2010PASP..122.1065P}, low-rank matrix decomposition for movie data \citep{Morii2017ApJ...835....1M}, discrete cosine transform \citep{Brueckner1995SoPh..162..357B, Belmon:Thesis:1998, Vohl:Thesis:2013}, and discrete wavelet transform \citep{Belmon2002A&A...386.1143B}. 

\begin{figure*}
\centering
\includegraphics[width=16cm]{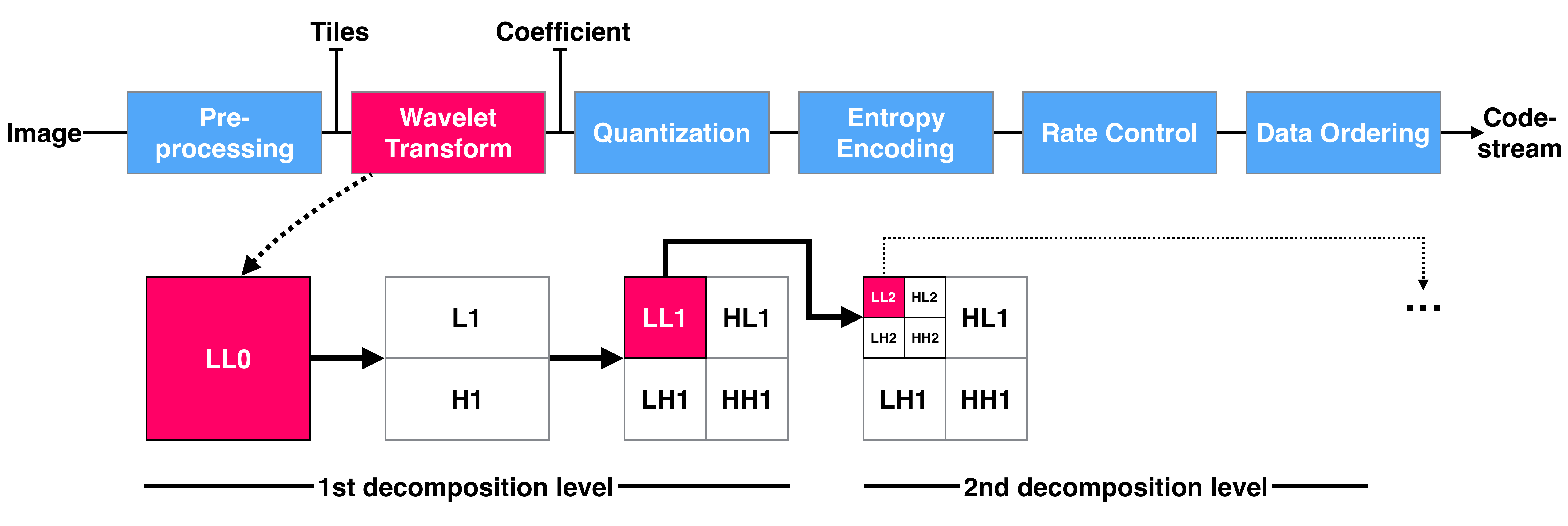}
\caption{JPEG2000 compression is applied as a stream of processing steps based on the discrete wavelet transform, scalar quantization, context modelling, entropy coding, and post-compression rate allocation [adapted from \citet{Kitaeff2015229}].}
\label{fig::jpeg2000encoding}
\end{figure*}

In this work, we consider the JPEG2000 (ISO/IEC 15444) standards \citep[part 1;][]{JPEG2000-part1} which offer lossy compression for both integer and real data. JPEG2000 compression is applied as a stream of processing steps that includes: pre-processing (tiling, level offset), wavelet transform\footnote{Lossy JPEG2000 implements the irreversible CDF-9/7 wavelet transform \citep{cohen1992biorthogonal}.}, quantization, entropy coding [via adaptive arithmetic coding \citep{Rissanen5390830}], rate control, and data ordering (Figure \ref{fig::jpeg2000encoding}).  A low level description of the standards, its algorithms and their related mathematics is beyond the scope of this paper.  Instead, we refer the reader to the JPEG2000 specification documentation and other related papers \citep[e.g.][]{JPEG2000-part1, Rabbani20023, li-2003}.  To evaluate the amount of storage space saved by compression, we use the concept of compression ratio.  We define the compression ratio ($\#$:1) as: 
\begin{equation}
\# = \frac{size_{o}}{size_{c}},
\label{eq::ratio}
\end{equation}
where $size_{o}$ is size of the original file and $size_{c}$ the size of the compressed file. 

Recent investigations of lossy JPEG2000 compression for astronomical images \citep{Peters-2014, Kitaeff2015229, Vohl2015200} show that it can lend high factors of compression while preserving scientifically important information in the data.  For example, \citet{Peters-2014} compressed synthetic radio astronomy data at several levels of compression, and evaluated how the loss affects the process of source finding.  In this case it was shown that the strongest sources (2000 mJy km/s and higher) could still be retrieved at extremely high compression ratio, where the compressed file would be more than 15,000 times smaller than the original file.  When using a high quantization step (compression ratio of about 90:1), low integrated flux sources (less than 800 mJy km/s) were still identified. 

To date however, no study has investigated the effect of lossy JPEG2000 on the process of transient finding, and no study has been conducted to evaluate its potential to accelerate data transmission in time-critical observation scenarios.  In this paper, we report on the evaluation of lossy JPEG2000 as part of DWF.

The remaining of the paper is structured as follows. Section \ref{sec::dwf-overview} presents a brief overview of the DWF science pipeline along with information about previous observation runs.  Section \ref{sec::custom-kerlumph} describes the compression software used for the experiments, and the rationale behind its custom design. Section \ref{sec::jpeg2000-pipeline} investigates the effect of lossy JPEG2000 on the DWF science pipeline.  In particular, Section \ref{sec::transient-search} presents the methodology and experimental results, evaluating the effect of compression on finding transient through the DWF science pipeline.  Section \ref{sec::timing} presents compression, decompression and transmission timing results obtained during DWF observation runs.  Finally, Section \ref{sec::discussion} discusses the results and their implications, while Section \ref{sec::conclusion} concludes and presents future work.

\section{Brief overview of the DWF science pipeline}
\label{sec::dwf-overview}

To date, DWF has seen a total of five observation runs, two pilot runs and three operational runs --- refining the overall practices each time.  The two pilot runs occurred during January and February 2015 respectively (pilot-1 and -2).  Since then, three operational runs occurred from 17--22 December 2015 UT (O1), 26 July to 7 August 2016 UT (O2), and  2--7 February 2017 UT (O3).  The grand lines of the science pipeline are as follows.  For detailed descriptions of the many DWF components, we refer the reader to \citet[][]{Meade2017arXiv170401281M, Andreoni-accepted}, and Cooke et al. (in prep.).

During the operational time of typical DWF run, three main steps are continuously being repeated for the optical data gathered by DECam:  

\begin{enumerate}
\item {\bf Data collection and transfer}
	\begin{enumerate}
	\item Images are acquired with DECam and saved as FITS files; 
	\item Each image is compressed to JPEG2000 and packaged to TAR;
	\item Each TAR is transferred to the Green II supercomputer;
    	\item Each TAR is unpacked, and each resulting image is decompressed.
	\end{enumerate}
\item {\bf Initial processing}
	\begin{enumerate}
	\item Individual CCD images are calibrated using parts of the {\tt PhotPipe} pipeline \citep{Rest2005ApJ...634.1103R};
	\item Image coaddition, alignment, and subtraction is performed using the {\tt Mary} pipeline \citep{Andreoni-accepted};
	\item {\tt Mary} generates a catalogue of possible transients, along with other data products (e.g. region files, small ``postage stamp'' images, light curves, etc.).
	\end{enumerate}
\item {\bf Visual inspection}
	\begin{enumerate}
	\item Visual analytics of potential candidates is performed by a group of experts and trained amateurs using an advanced visualisation facility \citep[see][]{Meade2017arXiv170401281M} and an online platform (database and other visualisation tools)\footnote{The online tools include: candidates list (ranked by priority), light curves, series of candidates ``cut-off'' images for visual inspection. The database and visual analytics tools are under the development of Sarah Hegarty (Swinburne University of Technology), Chuck Horst (San Diego State University) and collaborators.}.
	\item Provided that an interesting candidate is identified with sufficient confidence, a trigger is sent to the other telescopes for follow-up. 
	\end{enumerate}
\end{enumerate}

We note that steps 1b and 1c are executed in parallel, typically for about four files at a time on the observer's computer at CTIO.  Similarly, the step 1D, and the initial processing steps are executed in parallel for as many CCDs as possible on reserved computing nodes of the SwinSTAR\footnote{\url{http://supercomputing.swin.edu.au/about-green-ii/}} component of the Green II supercomputer.

\section{Software design rationale}
\label{sec::custom-kerlumph}
In the time-critical scenario of DWF, a gain in transmission time offered by data compression is only interesting if compression and decompression can be achieved quickly.  To this end, \citet{Vohl2015200} demonstrated that \texttt{KERLUMPH}\footnote{\url{http://supercomputing.swin.edu.au/projects/kerlumph/}} --- a multi-threaded implementation of the JPEG2000 standard --- can compress and decompress large files quickly.  

For a sample of 1224 files, \texttt{KERLUMPH} achieved both compression and decompression of a $400$ megabyte (MB) binary file in less than 10 seconds, with median and mean time under 3 seconds using the Green II supercomputer. The tests on Green II were performed using Linux (CentOS release 6.6) running on SGI C2110G-RP5 nodes -- one node at a time -- containing 2 eight-core SandyBridge processors at 2.2 GHz, where each processor is 64-bit 95W Intel Xeon E5-2660.

\begin{figure}
\centering
\includegraphics[width=8cm]{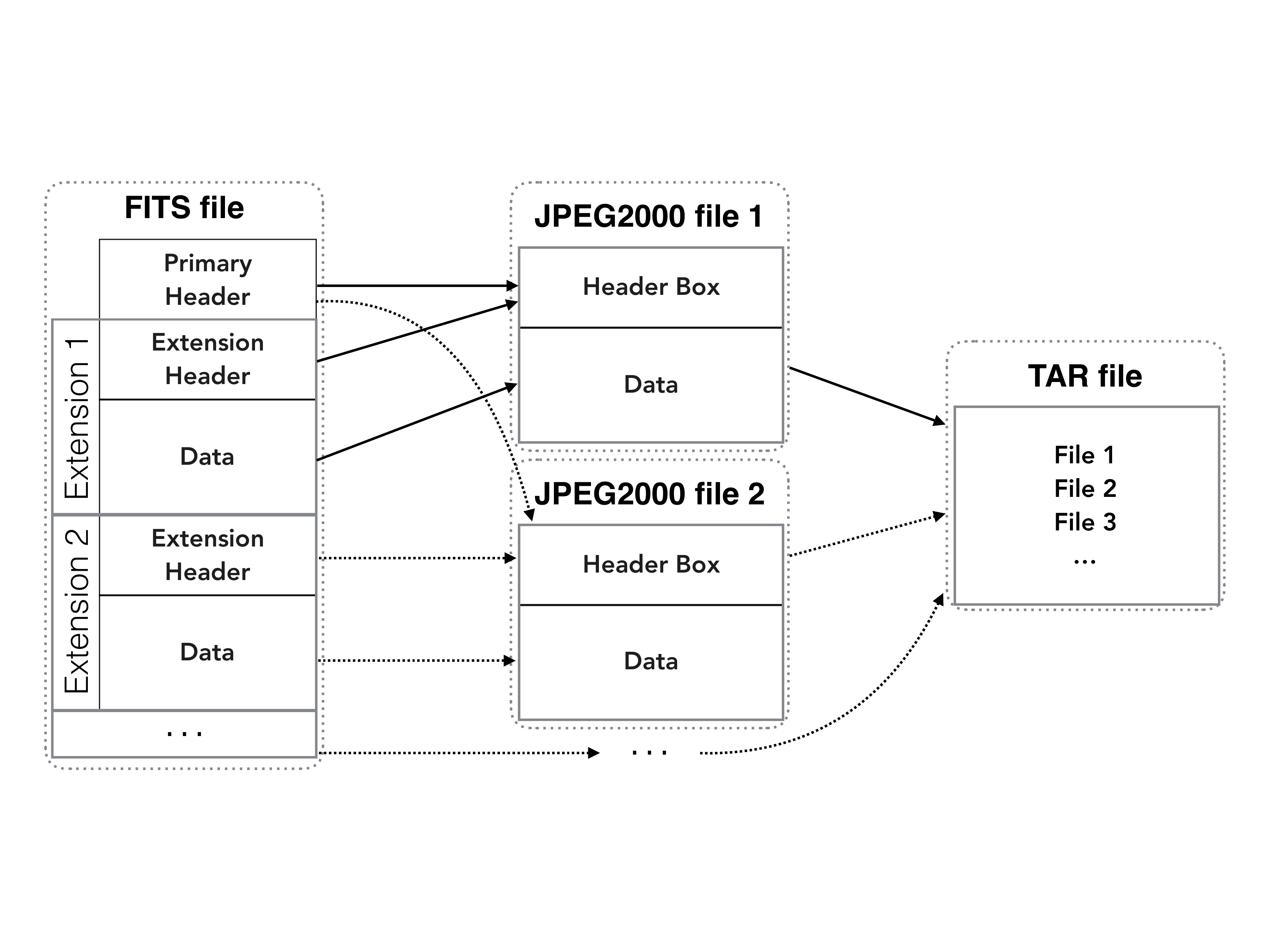}
\caption{Compression procedure schematic diagram. The multi-extension FITS file from DECam is lossily compressed into multiple JPEG2000 (one per extension), and then grouped together into a TAR file ready for transmission.  Note that the primary header is merged with the extension header.}
\label{fig::compress}
\end{figure}

We modified \texttt{KERLUMPH} to specifically compress the FITS files from DECam into JPEG2000.  In addition of allowing the compression of FITS files, we customized the compression pathway to modify the input file in a number of ways (Figure \ref{fig::compress}).  At compression, the multi-extension FITS file of DECam is lossily compressed into multiple JPEG2000 files (one per extension) --- merging its specific Extension Header with the Primary Header.  The rationale behind this decision is 
to simplify parallel processing in the next steps of the pipeline. 

To avoid having to send $\sim60$ individual files over the internet, we group them together into a single TAR file\footnote{TAR is an archive format that collects any number of files, directories, and other file system objects into a single stream of bytes. See \url{https://www.gnu.org/software/tar/} for more details.} before transmission.  To save extra space, we do not include information relative to guide CCDs.  Moreover, at the time of observation, two CCDs (at position S2 and N30) were not working and the amplifier of another CCD (S26) had a defect leading to difficult calibration.  The cumulated raw data of these CCDs represents $\sim100$ MB that would need to be compressed and transmitted, to be eventually left out of the analysis.  We therefore decided to discard these extensions for the near real-time analysis.

\begin{figure}
\centering
\includegraphics[width=8cm]{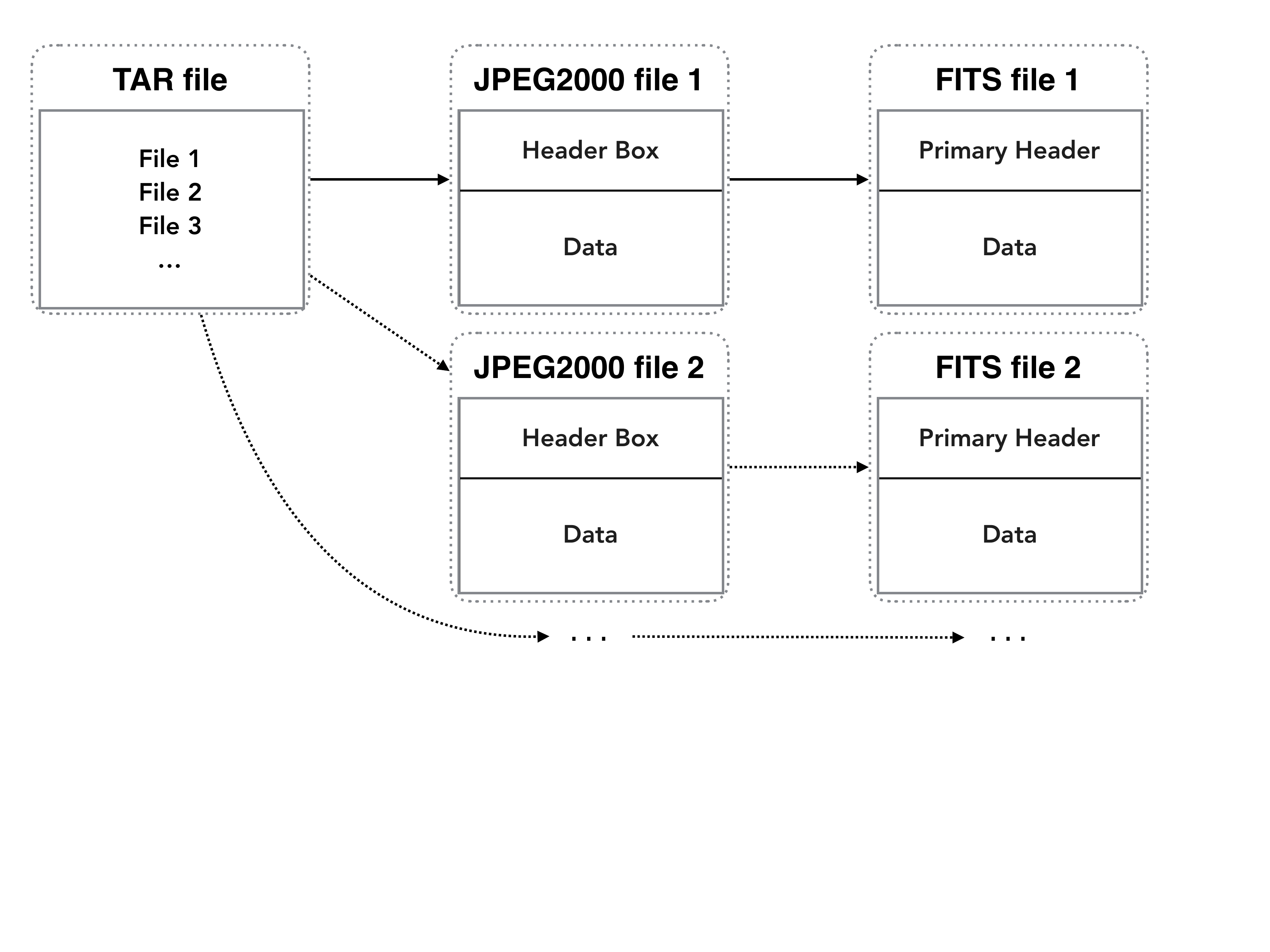}
\caption{Decompression procedure schematic diagram. The TAR file is expanded to recover all JPEG2000 files; each of them is then decompressed into a single extension FITS file.  Each FITS file corresponds to a given extension of the original file, where the primary header contains the merged information of the original primary header and the current extension header.}
\label{fig::decompress}
\end{figure}

At decompression (Figure \ref{fig::decompress}), the software recreates the FITS file using the \texttt{cfitsio} library --- as several of the subsequent processing steps, many using standard ``off-the-shelf'' available tools, do not yet support JPEG2000.  The recreation of the FITS file enables us to proceed with the pre-processing required by {\tt PhotPipe}.  During this phase, we add and modify specific keywords in the header, avoiding the slow procedure of updating the FITS header further down in the pipeline (see Appendix \ref{sec::appendix1} for more details). 

Similar to the original version of {\tt KERLUMPH}, the modified version allows setting and modifying JPEG2000 compression parameters (Part 1, shown in Table \ref{table::jpeg2000params}).  This capability includes the coefficient quantization step size ({\em Qstep}) --- which is used to discretise the wavelet coefficient values --- and the number of wavelet transform levels ({\em Clevels}) used to influence the wavelet domain before quantization and encoding \citep{Clark2008}.  In addition, it is possible to specify a target bit-rate parameter ({\em rate}) to set an upper limit on the output storage size.  This is done via the post-compression rate allocation, in which the compressed blocks are passed over to the Rate Control unit.  The unit determines how many bits of the embedded bit-stream of each block should be truncated to achieve the target bit rate --- aiming to minimise distortion while still reaching the target bit-rate \citep{Kitaeff2015229}. 

\begin{table}
  \centering
    \caption{Parameters in Part 1 of the JPEG2000 Standard, ordered as encountered in the encoder.  The only parameter for which the default value is modified during an observation run is highlighted.}
    \begin{tabular}{cl}
        \hline
        & Parameter  \\
        \hline
1. & Reconstructed image bit depth \\
2. & Tile size \\
3. & Color space \\
4. & Reversible or irreversible transform \\
5. & Number of wavelet transform levels \\ 
6. & Precinct size \\
7. & Code-block size \\
{\bf 8.} & {\bf Coefficient quantization step size} \\
9. & Perceptual weights \\
10. & Block coding parameters: \\
&(a) Magnitude refinement coding method \\
&(b) MQ code termination method \\
11. & Progression order \\
12. & Number of quality layers \\
13. & Region of interest coding method \\
     \hline
    \end{tabular}
  \label{table::jpeg2000params}
\end{table}

\citet{Peters-2014} show that the code block size and precincts size had no effect on both compression and soundness of their spectral cube data.  Therefore, we have bypassed these parameters for this evaluation. 
\citet{Vohl2015200} show that the combined use of {\em Qstep} and a high {\em Clevels} value can increase the compression ratio while preserving a similar root-mean-squared-error, as the wavelet decomposition levels increase for a similar quantization step size.  However, we do not increase {\em Clevels} from the default value of 5 in the context of DWF.  An increased {\em Clevels} value requires a larger amount of random access memory --- as more level of wavelet decomposition are being processed --- which would penalise us while we aim to reduce the weight of the compression on the overall computation at CTIO.

\section{Effect of lossy JPEG2000 on the DWF science pipeline}
\label{sec::jpeg2000-pipeline} 

In this Section, we evaluate the effect of using lossy JPEG2000 as part of the DWF science pipeline.  In particular, we present an experiment evaluating how the different levels of compression affect the process of transient finding with the DWF science pipeline.  Finally, we report on transmission time recorded during the O2 and O3 run. 

\subsection{Effect on transient search}
\label{sec::transient-search}

While we note that all raw data for DWF is archived and can be evaluated at a later time, it is nevertheless important to evaluate how lossy JPEG2000 affects the process of finding transients for the near real-time analysis.  As DWF uses a custom pipeline, 
we use it integrally in this experiment. We refer the reader to \citet{Andreoni-accepted} for details on the {\tt Mary} pipeline, based on image subtraction techniques, and its candidate selection parameters.  
Furthermore, to provide a realistic case study (e.g. instrumental noise characteristics), we use raw images obtained with DECam during the DWF O2 run as the starting point of the experiment.
The results of this study finds no significant loss of transient detection at all brightnesses relevant to the DWF survey to compression ratios $\sim$25:1. 

\subsubsection{Methodology}
\label{sec::transient-search-methodology}

We select three raw FITS images from DECam obtained on 2016-08-02, between 09:22:05 UTC and 09:42:08 UTC (post exposure time).  While it would be possible to identify transients within these images directly, it would also be a difficult task to assess their reliability and intrinsic parameters --- a task that we reserve for future DWF papers.  Instead, we manually inject artificial transient sources for which we know the characteristics in advance (e.g. flux, position, point spread function). We set the range of injected sources between magnitudes 17 (brightest) to 26 (faintest) to probe the detection limits of the survey --- which is expected to have a minimum source detection magnitude of $\sim$22.3--22.5 for these images. 

\begin{figure*}[!ht]
\centering
\includegraphics[width=16cm]{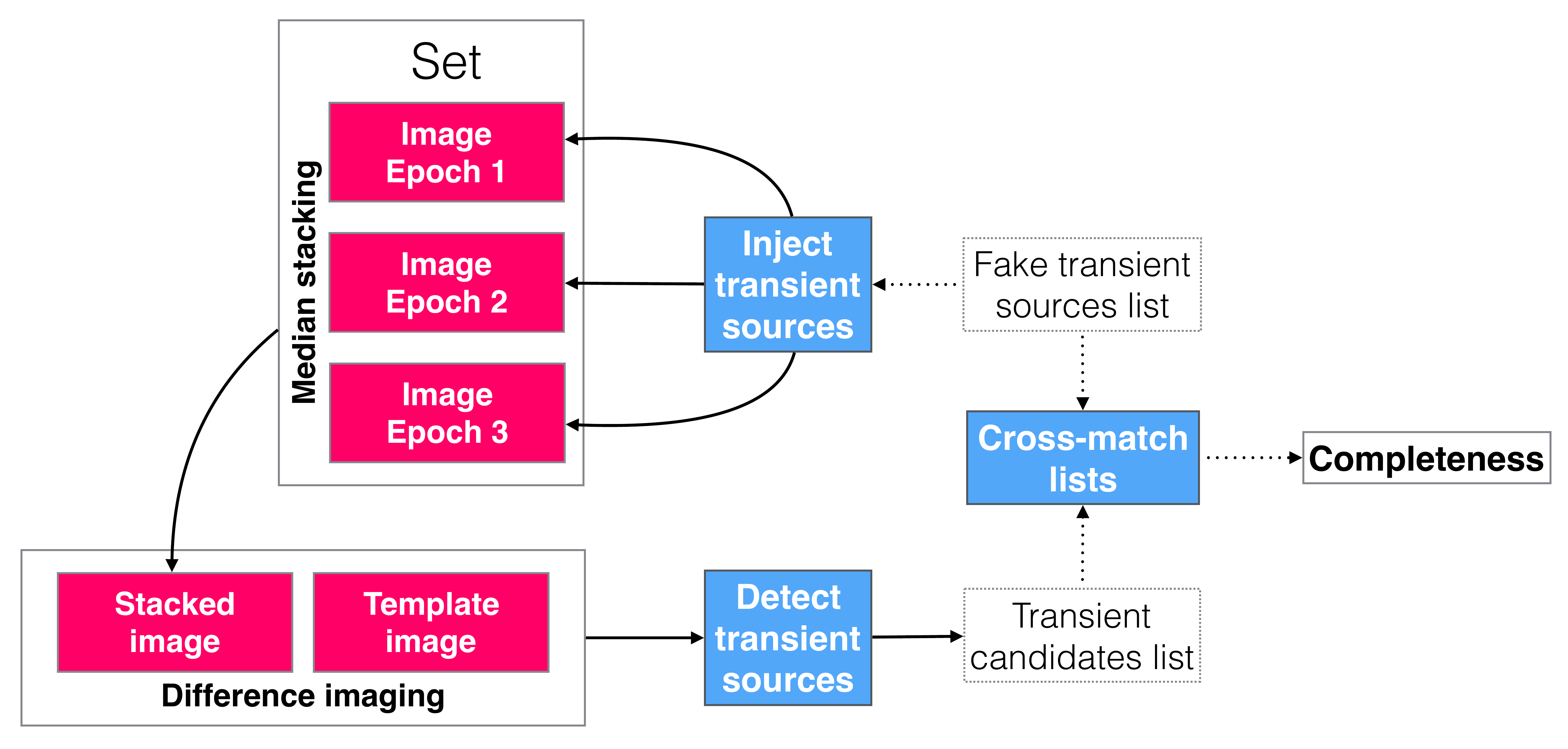}
\caption{Schematic diagram of the experiment setup.  Three images taken at different epochs form a set.  Transients are added to each image of the set, using the same sky coordinates in all three images. Images of the set are coadded (median stacking) to better detect the transients, and to eliminate cosmic rays --- reflecting a transient lasting longer than three images worth of time (about 120s). Difference imaging is then applied between the stacked image and a template image, resulting in a residual image.   Transient detection is applied on the residual image using the {\tt Mary} pipeline, which outputs a transient candidates list.  This list is cross-matched with the list of injected sources to evaluate the completeness. We note that a loss in completeness will natually occur when sources fall onto bright sources, making its detection difficult or impossible.  
}
\label{fig::source-finding}
\end{figure*} 

All three images are used as a set for transient detection.  In addition, an image taken on 2016-07-28 and processed using the DECam Community Pipeline  \citep{Valdes2014ASPC..485..379V} is used as template.  Transients are added to every image in the set.  For each CCD in the set, we inject 273 sources (2D gaussian) drawn from a uniform distribution of magnitudes.  The range is split into bins of 0.1 magnitude, corresponding to 3 sources per magnitude bin.  Sources locations are allocated randomly, while avoiding a 75 pixel border around the edge of the CCD --- to avoid being cropped out during the alignment process.  Sky coordinates are preserved throughout all images (e.g. a given source is found at the same location in all images). Sources are generated using the {\tt make\_gaussian\_sources} function of the {\tt photutils} package \citep{larry_bradley_2016_155353}, an affiliated package of {\tt astropy} \citep{Astropy2013A&A...558A..33A}.

Each image is compressed at several fixed compression ratio, ranging from 5:1 to 100:1, with a step of 5. To do so, we set the {\em rate} parameter to the ratio between the original {\tt BITPIX} value of the 32-bit image to the desired compression ratio ($D_{\#} \in [5, 10, 15, ..., 90, 95, 100]$): 
\begin{equation}
rate = \frac{{\tt BITPIX}}{D_{\#}}.
\label{eq::rate}
\end{equation}
 
For each level of compression, we proceed with the {\em initial processing} steps of the pipeline (Section \ref{sec::dwf-overview}). 
The three images in the set are calibrated, aligned, and coadded (image stacking) to better detect the transients, and to eliminate cosmic rays --- reflecting a transient lasting longer than three images worth of time (about 120s).  The stacked image is used for difference imaging with the template image.  Finally, we cross-match the {\tt Mary} pipeline's candidate list with the list of positions for the injected sources. In this context, we define the transient finding completeness $c_{\#}$ for compression ratio $\#$:1 as:
\begin{equation}
c_{\#} = \frac{N_{M,\#}}{N_{I}},
\label{eq::completeness}
\end{equation}
where $N_{M,\#}$ is the number of sources found by {\tt Mary} for a file compressed at a ratio of \#:1, and $N_{I}$ is the number of transients injected.  We normalize $c_{\#}$ by comparing it to the completeness obtained with the original data $c_{1}$ (never compressed) to avoid reporting biases incoming from the source finder that are unrelated to this work.  Therefore, we report the normalized completeness $\mathcal{C}_{\#}$ for compression ratio \#:1 as: 
\begin{equation}
\mathcal{C}_{\#} = \frac{c_{\#}}{c_{1}}.
\label{eq::norm_completeness}
\end{equation}
An overview of the different steps of the experiment is shown in Figure \ref{fig::source-finding}. 

\subsubsection{Results}
\label{sec::transient-search-results}

\begin{figure*}[!ht]
\centering
\includegraphics[width=17.3cm]{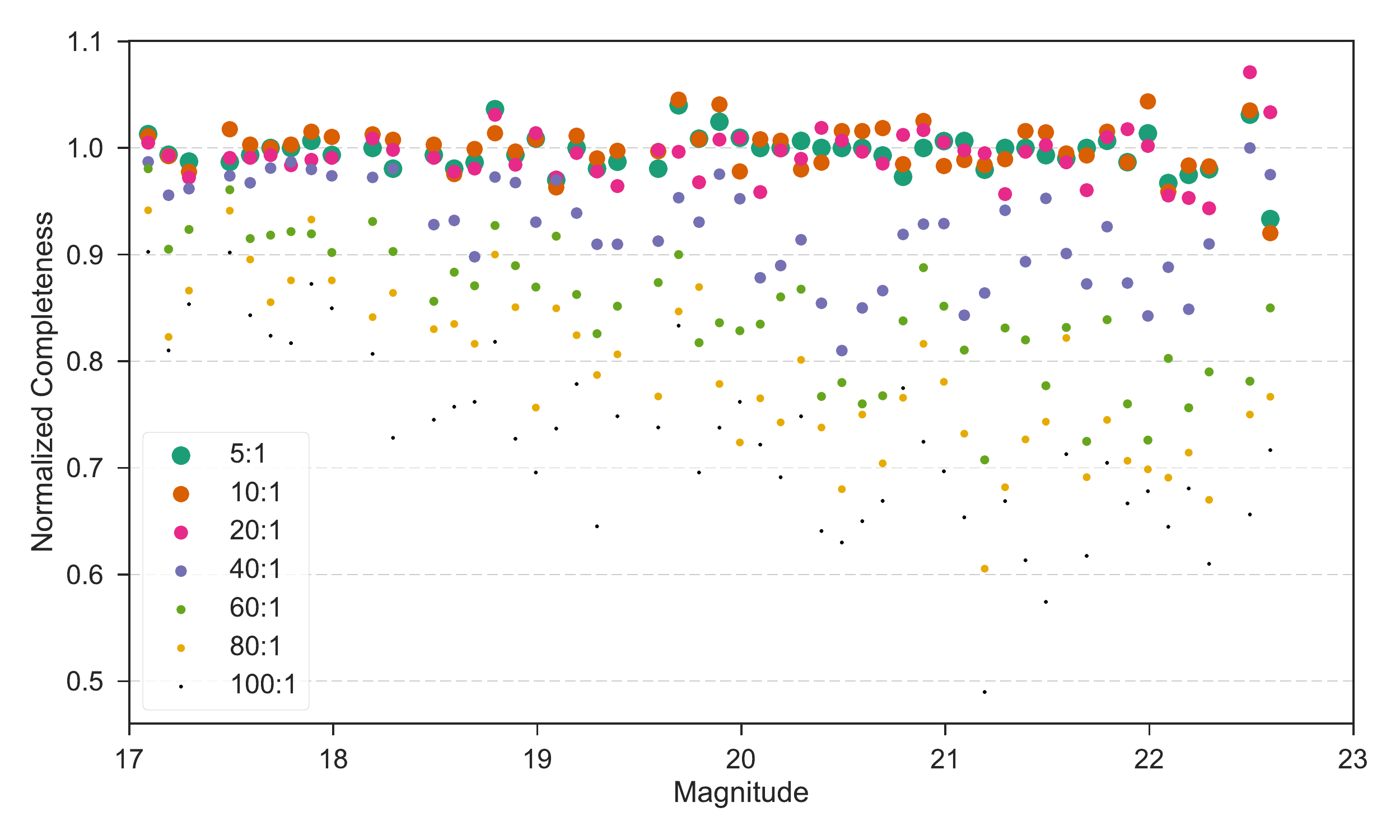}
\caption{Normalized completeness as a function of magnitude for all evaluated compression ratio (Equation \ref{eq::norm_completeness}). A normalized completeness of 10$^{0}$ indicates no difference in transient finding results between the compressed and never-compressed data. Results above and below this line show that compression affected the findings positively or negatively respectively. Results are limited to cases where completeness on original data was $\geq$0.5 (i.e., down to magnitude $\sim$ 22.5).}
\label{fig::completeness}
\end{figure*}

Figure \ref{fig::completeness} shows the normalized completeness as a function of magnitude for the different compression ratios.  The four panels split the compression ratios into groups of five (i.e. the first panel shows results for compression between 5:1 and 25:1 inclusively, the second panel shows results between 30:1 to 50:1, and so on).  Results are limited to cases where a completeness $\geq$0.5 was found by the {\tt Mary} pipeline for the original data (never compressed) --- which eliminates data below our detection threshold (i.e., down to source magnitudes of $\sim$22.5).

A normalized completeness of 1.0 indicates that compression had no effect on the process of finding transients compared to working with original data (never compressed).  Results above and below this line show that compression affected the findings positively (more transients were correctly identified) or negatively (less transients were identified), respectively.

As expected, as the compression ratio increases, the number of sources missed by the source finder also increases.  In general, fainter sources are more affected by compression than brighter sources, while the brightest sources are the least affected overall.  This is noticeable when comparing the slopes of the distributions,   increasing in steepness as the compression ratio increases.  We find that compression up to about $\sim$25:1 has a negligible effect on the process of finding transients, and only a small affect for the faintest magnitude.  This result can be further confirmed by looking at the mean normalized completeness.

\begin{figure*}[!ht]
\centering
\includegraphics[width=17.3cm]{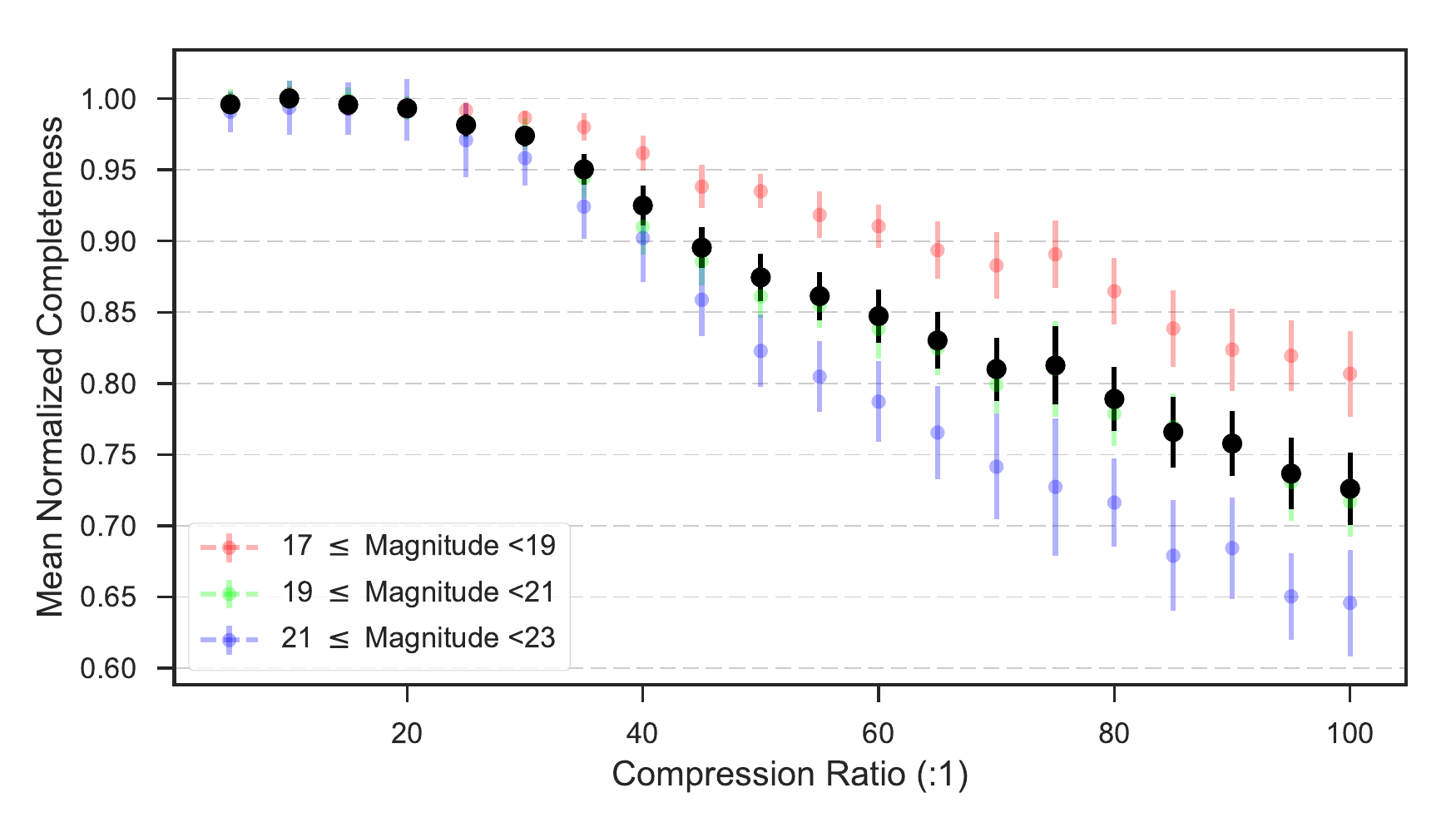}
\caption{Mean normalized completeness and 95\% confidence interval as a function of compression ratio. Results are limited to cases where completeness on original data was $\geq$0.5. 
}
\label{fig::mean-completeness}
\end{figure*}

Figure \ref{fig::mean-completeness} shows the mean normalized completeness as a function of compression ratio.  In addition, the error bars indicate the 95\% confidence interval, defined as: 
\begin{equation}
\epsilon = \frac{2\sigma}{\sqrt{N}},
\label{eq::interval}
\end{equation}
where $\sigma$ is the standard deviation, and $N$ is the number of sources used to evaluate the normalized completeness.  The black markers show the overall mean value per compression ratio. Results show that a compression up to 35:1 provides on average a normalized completenesses $\geq$ 95\%, and $\geq$ 90\% for a compression ratio up to 40:1. 

Figure \ref{fig::mean-completeness} also shows the mean normalized completeness for three magnitudes ranges.  Specifically, the red markers indicate the mean for magnitudes between 17 and 19, the green markers for magnitudes between 19 and 21, and the blue markers for magnitudes between 21 and 23.  Breaking down magnitude range this way highlights how bright sources (mag = 17--19) are less affected by compression than the faintest sources (mag = 21--23), where the mean normalized completeness decreases faster for fainter sources.  In all cases however, results show that a relatively high compression ratio of 30:1 has minimal impact on source finding, where sources of magnitudes between 21 to 23 show a mean normalized completeness >95\% in the DECam images as compared to source identification in non-compressed data. 

Another concern is that the time savings gained due to the usage of lossy compression may be diminished if a significant number of false positives sources are detected (requiring human validation) compared to our fiducial baseline.  From this experiment, we find the total number of identified sources to be within $\lesssim$5--8\% of those found without compression at compression ratios of $\lesssim$35:1, and an increase up to $<$20\% at higher compression ratios. Furthermore, during an observation run, the behaviour of the transients will further `clean' the data of any false positive detection from compression. DWF only triggers other telescopes on transient sources with $\sim$30 minutes to hours duration. Therefore, a transient must be detected in multiple images (more than three images) to be considered a true candidate by the campaign.

From these results, we estimate that utilising lossy JPEG2000 compression with a compression ratio up to 25:1 enables the DWF team to efficiently retrieve transient sources within the detection limits of the survey without significant loss.  These results are in agreement with those obtained by \citet{Peters-2014} 

\subsection{Timing}
\label{sec::timing}
In this Section, we evaluate how compression accelerates data transmission from CTIO (Chile) to the Green II supercomputer (Australia).  We evaluate the speed-up factor in data transfer time, in addition to compression and decompression time.  Timing data was recorded for a total of 13,081 files during the O2 (year 2016) and O3 (year 2017) observation runs.  We find that the speed-up factor in transfer time outweigh the compression and decompression time --- validating the decision of integrating lossy data compression as part of our pipeline.

\subsubsection{Estimation of the data transfer acceleration}
\label{sec::Transfer-timing}
For each file transferred (using the unix command {\tt scp}), we record the size of the compressed file and the transfer time.  From these two measures, we evaluate the compression ratio as defined by Equation \ref{eq::ratio}, where $size_{o}$ is equal to 1184 MB.  We also evaluate the transfer rate, defined as:
\begin{equation}
r = \frac{size_{c}}{t},
\label{eq::transfer_rate}
\end{equation} 
where $size_{c}$ is the size of the compressed file (MB), $t$ is the transfer time (s) of the compressed file, and $r$ is the transfer rate (MB/s). The compression ratio, transfer time, and transfer rate distributions for each day of the O2 and O3 observation runs are provided in Appendix \ref{fig::rate-v-date}. Compression ratio was varied by the team during each run to provide data with visual quality as high as possible, while providing fast enough transfer time. As the transfer rate varies during an observation run, we proceed with the following method to evaluate the speed-up factor provided by compression.

\begin{figure*}[!ht]
\centering
\includegraphics[width=17.3cm]{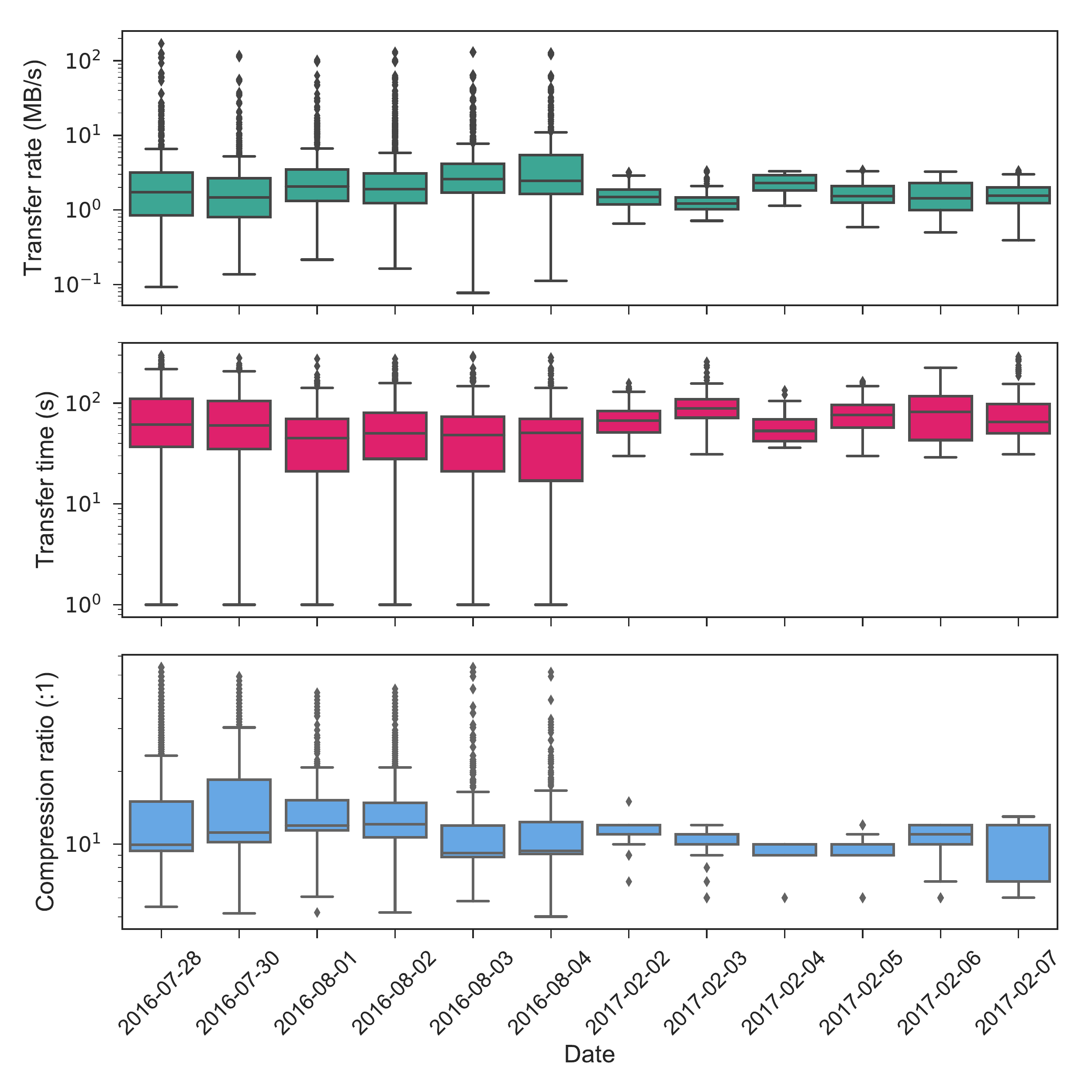}
\caption{Box and whiskers plot showing distributions of transfer rate (MB/s, top panel), transfer time (s, central panel) and compression ratio (:1, bottom panel) obtained during each day of the O2 (2016) and O3 (2017) runs. The median (line) is within the box bounded by the first and third quartiles range ($\mathtt{IQR}$ = $\mathtt{Q3}-\mathtt{Q1}$). The whiskers are $\mathtt{Q1-1.5 \times IQR}$ and $\mathtt{Q3+1.5 \times IQR}$. Beyond the whiskers, values are considered outliers and are plotted as diamonds. We note that transfer rate varied greatly from day to day. Compression ratio was varied by the team during each run to provide data with visual quality as high as possible, while providing fast enough transfer time.}
\label{fig::rate-v-date}
\end{figure*}

For each transferred file, we estimate the transfer time $\hat{t}$ (s) that would have been required without compression, assuming the transfer rate at the time of transmission: 
\begin{equation}
\hat{t} = \frac{size_{o}}{r}.
\label{eq::estimated_transfer_time}
\end{equation} 
Using $\hat{t}$, we estimate the speed-up factor $\hat{s}$ for a given file, defined as:
\begin{equation}
\hat{s} = \frac{\hat{t}}{t},
\label{eq::estimated-speed-up}
\end{equation}
Similarly, we estimate the saved transmission time $\hat{\theta}$ (s) for this file: 
\begin{equation}
\hat{\theta} = \hat{t} - t.
\label{eq::estimated-time-saved}
\end{equation}

\begin{figure*}[!ht]
\centering
\includegraphics[width=17.3cm]{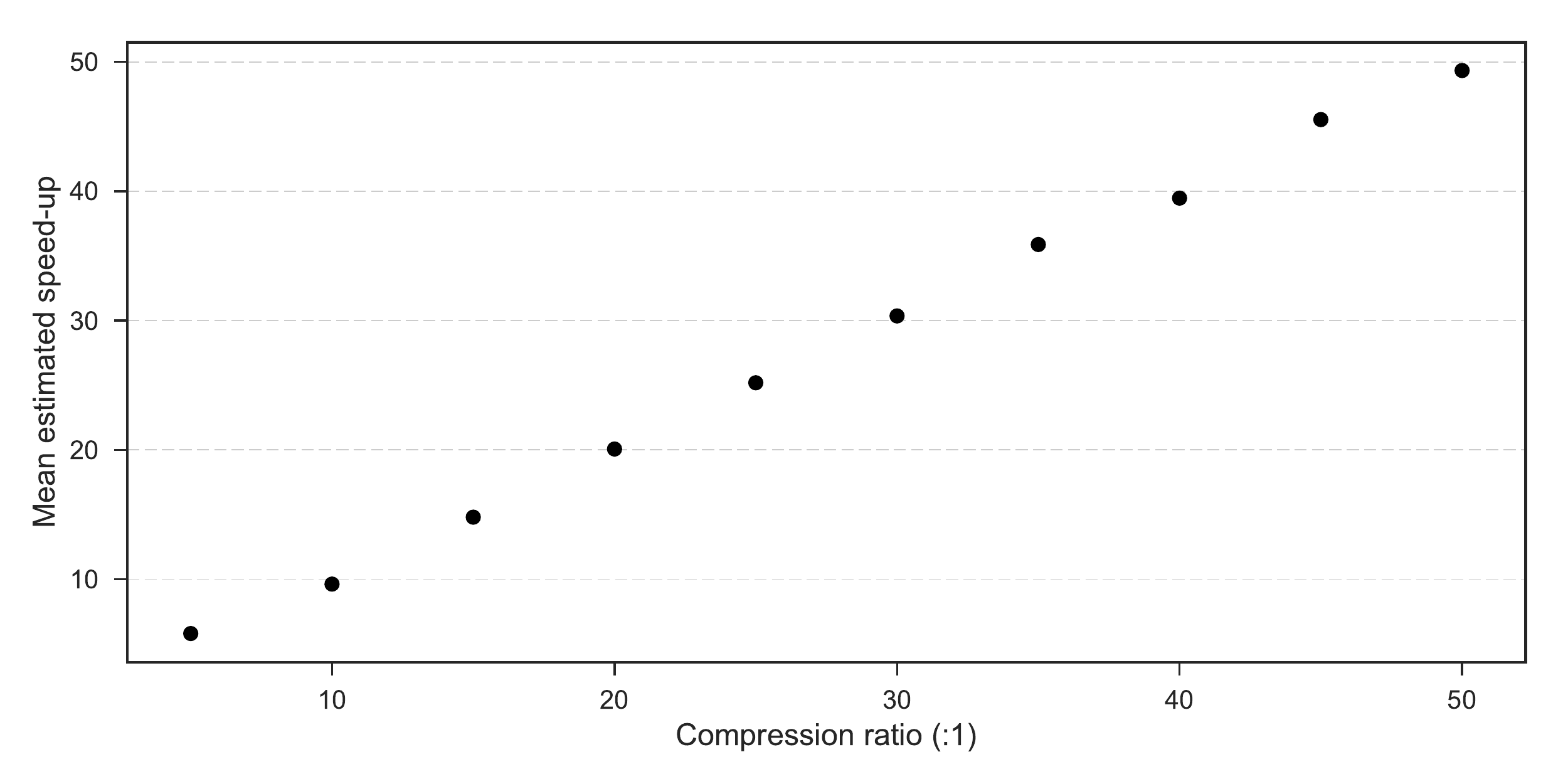}
\caption{Mean estimated speed-up factor ($\hat{s}$) and 95\% confidence interval as a function of compression ratio for 13,081 files transferred during the O2 run (2016) and O3 run (2017). }
\label{fig::mean-speedup}
\end{figure*}

Figure \ref{fig::mean-speedup} shows the mean estimated speed-up factor $\hat{s}$ and 95\% confidence interval (Equation \ref{eq::interval}) as a function of compression ratio.  The summary of results obtained during the O2 and O3 runs are shown in Table \ref{table::timing-results}.  From these results, we note a linear relation between compression ratio and the estimated speed-up factor. During the two observation runs evaluated (O2 and O3), we obtained a mean compression ratio of 13:1 (targeted 'on-the-fly' by the team), which provided a mean estimated speed-up factor of 13.04 --- equivalent to an estimated 14.60 minutes saved per file transfer. 

\begin{table*}[!ht]
\centering
\caption{Summary of transmission timing results for the combined (Both) and individual (O2, O3) observation runs.  Columns show compression ratio (\#), transfer rate ($r$), estimated speed-up factor ($\hat{s}$), and estimated transfer time saved ($\hat{\theta}$).  Rows show minimum, maximum, mean, median, and standard deviation of the distribution.}
\label{table::timing-results}
\begin{tabular}{lllllllllllll}
\hline
\hline
\textbf{} & \multicolumn{3}{l}{\#\textbf{:1}}         & \multicolumn{3}{l}{\textbf{$r$ (MB/s)}}   & \multicolumn{3}{l}{\textbf{$\hat{s}$}}          & \multicolumn{3}{l}{\textbf{$\hat{\theta}$ (minute)}}            \\
\textbf{} & \textit{Both} & \textit{O2} & \textit{O3} & \textit{Both} & \textit{O2} & \textit{O3} & \textit{Both}   & \textit{O2}     & \textit{O3} & \textit{Both}      & \textit{O2}        & \textit{O3}        \\
\hline
\textit{Min} & 5.02 & 5.02 & 6.00 & 0.08 & 0.08 & 0.39 & 5.02 & 5.02 & 6.01 & 0.10 & 0.10 & 4.76 \\
\textit{Max} & 53.82 & 53.82 & 15.00 & 170.00 & 170.00 & 3.46 & 53.82 & 53.82 & 15.58 & 250.01 & 250.01 & 46.14 \\
\textit{Mean} & 13.00 & 13.18 & 10.45 & 5.09 & 5.32 & 1.61 & 13.04 & 13.18 & 11.01 & 14.60 & 14.71 & 13.02 \\
\textit{Median} & 10.21 & 10.12 & 11.00 & 2.05 & 2.11 & 1.44 & 10.21 & 10.12 & 11.72 & 8.74 & 8.41 & 12.43 \\
\textit{Std} & 7.26 & 7.46 & 1.80 & 13.56 & 13.97 & 0.66 & 7.25 & 7.46 & 1.82 & 20.93 & 21.58 & 5.26 \\
\hline
\hline
\end{tabular}
\end{table*}

\subsubsection{Compression and decompression time}
We also recorded the time required to compress the data at CTIO during both runs.  Compression was performed on the observer computer at CTIO.  The computer includes an ASUS P6X58D LGA 1366 motherboard with 24 GB of DDR3 1600 memory, an i7-950 quad core processor in the LGA 1366 form factor, and 3 TB of hard disk drive (HDD) for storage.  We note that compression is only one of many processes running on the observer computer --- where it is common to have multiple internet browser windows opened onto SISPI\footnote{\url{https://des.mps.ohio-state.edu/Tools/sispi_main.htm}} (the DECam software), weather stations, etc., in addition to any other software used by the observer.  We obtain a mean, median, minimum and maximum compression time of 42.49, 37.75, 33.27 and 84.81 seconds respectively, with a standard deviation of 9.95 seconds. 

Decompression is performed on the Green II supercomputer.  Contrary to the experiment performed by \citet{Vohl2015200} --- which proceeded with decompression on the Lustre File System\footnote{Lustre File System, [online] Available: http://www.lustre.org.} directly --- we perform the decompression via the local storage of Green II (using {\tt PBS\_JOBFS}) to obtain fast read and write access to HDD storage.  We obtain a mean and median of 1.82 seconds, minimum of 1.57 seconds, and maximum of 1.86 seconds, with a standard deviation of 0.03 seconds.  This timing represent the decompression of a single CCD. 
Therefore, one needs to cumulate the time for all the 57 CCDs.  However, as this is performed in parallel on Green II , this cumulated time does not reflect user wait time.  

\section{Discussion}
\label{sec::discussion}

For projects dealing with very large datasets, a perfect scenario would be that all data processing would be done on site at the data acquisition location, with minimal data movement. However, as it is still common for international teams to post-process their data on  local computing and supercomputing resources, the need to transfer data is unlikely to be removed completely on short time-scale.  In this context, lossy compression provides faster data transfer to execute science otherwise not possible in fast targeted time-scales, including the near real-time data processing required for the DWF survey.

While the addition of lossy compression to the science pipeline of DWF introduces an additional need for care by the team, the discovery of potential transients, including supernova shock breakouts, off-axis GRBs, counterparts to FRBs and gravitational waves, and other highly sought-after sources, along with flare stars, cataclysmic variables, x-ray binaries, etc., highlight its ability to accelerate discovery in time-critical scenarios. 

As the steps between data acquisition and transient confirmation are dependent on one another, the transfer time speed-up factor provided by data compression reduces the overall time before a trigger can be sent to other observatories.  Further comparative investigation of the process of transient finding --- comparing results obtained with and without lossy compression --- should provide insights on the necessity of using raw data.

During run O2 and O3, file transfer to Australia was faster than the rate in which the data could be processed using the version of the reduction pipeline used at that time. Hence, the compression ratio was manually set by a member of the observatory team at CTIO, using the {\em Qstep} parameter in order to regulate transfer time.  Regulation considered current empirical internet speeds and data processing status in Australia.  Doing so, creates data with the highest visual quality as possible, while providing the necessary fast transfer time.  During O2 and O3, the team aimed for transfer time $\sim$1--2 minutes.  A compression ratio $\leq$ 20:1 was judged to be a comfortable upper limit for transfer time, and a safe choice in term of loss and visual quality. 

The timing results show however that transfer rates can vary significantly during an observation run, and hence, the compression ratio is not the only factor that influences the total transfer time.  Future work should evaluate methods to automatize the compression parameters selection (e.g. {\em Qstep}, {\em CLevels}) to provide the minimal loss for a targeted bit rate selected for a targeted optimal transfer time (based on criteria defined by the team).  Future investigation to further accelerate data transmission should consider tracking individual packet transmission to identify bottlenecks. 


\section{Conclusions and future work}
\label{sec::conclusion}
When considering the three {\em V}s of ``Big Data'' (variety, volume, velocity), volume and velocity have a direct impact on modern astronomy endeavours, such as time domain science.  In recent years, the DWF initiative --- a collaborative, remote and time-critical observation program --- has been detecting and identifying, in near real-time, fast transient events on millisecond-to-hours timescales using DECam and $\sim$20 other telescopes.  Data files generated with DECam are large ($\sim 1.2 GB$ per FITS file) and the high volume of short-exposure images provide data of imposing size.

To search for transient candidates in near real-time imposes computational requirements exceeding the processing capacity available on-site at the observatory in Chile.  Instead, data files need to be constantly transmitted to the Green II supercomputer in Australia for post-processing, source finding, visualization and analysis.  To reduce the stress imposed by the transmission of large amount of raw data, we integrate lossy data compression as part of the science pipeline --- keeping in mind that raw data is archived and can be evaluated at a later time. 

In this paper, we present an evaluation of the impact of lossy JPEG2000 on the DWF pipeline.  In particular, we estimate that the compression ratio is linearly related to the speed-up factor.  In particular, the average measured file compression ratio of $\sim$13:1 during two DWF observation runs, resulted in a mean estimated speed-up factor of 13.04.  In addition, we find that the speed-up factor outweighs the added compression and decompression time.  

We also presented an experiment evaluating the impact of lossy JPEG2000 on the process of finding transient sources.  We find that utilising compression ratios up to 30:1 will enable transient source detection to the detection limits of the survey with negligible efficiency losses, and $\sim$10--15 minutes saved per file transfer --- enabling rapid transient science that would otherwise not be possible.  These results validate the choice of integrating lossy data compression to accelerate the overall DWF scientific pipeline.

\begin{acknowledgements}
This research was undertaken with the assistance of resources provided by the Green II supercomputer through the ASTAC scheme supported by the Australian Government's Education Investment Fund.  We thank Tim Abbott for providing the hardware information for the observer's computer at CTIO.  DV thanks George Bekiaris, Christopher J. Fluke and Amr H. Hassan for valuable discussions during the development of the KERLUMPH software. 
Research support to IA is provided by the Australian Astronomical Observatory. JC acknowledges support of this research by the ARC Future Fellowship grant FT130101219. Parts of this research were conducted by the Australian Research Council Centre of Excellence for All-sky Astrophysics (CAASTRO), through project number CE110001020. Parts of this research were conducted by the Australian Research Council Centre of Excellence for Gravitational Wave Discovery (OzGrav), through project number CE170100004.

\end{acknowledgements}

\begin{appendix}
\section{Details of FITS header modifications}
\label{sec::appendix1}

We modify the \texttt{OBSTYPE} value following Equation \ref{eq::update-OBSTYPE}, and add three keywords based on the current header's content: \texttt{RDNOISE}, \texttt{GAIN}, and \texttt{SATURATE}. The definitions of these keywords are expressed in Equations \ref{eq::add-RDNOISE}, \ref{eq::add-GAIN}, and \ref{eq::add-SATURATE}.

\begin{equation}
\texttt{OBSTYPE} = \left\{ \begin{array}{rl}
 \text{bias}, &\mbox{\text{if } \texttt{OBSTYPE} = \text{zero}} \\
 \text{domeflat}, &\mbox{\text{if } \texttt{OBSTYPE} = \text{dome flat}}
\end{array} \right.
\label{eq::update-OBSTYPE}
\end{equation}

\begin{equation}
\texttt{RDNOISE} = \frac{1}{2}\times (\texttt{RDNOISEA} + \texttt{RDNOISEB})
\label{eq::add-RDNOISE}
\end{equation}

\begin{equation}
\texttt{GAIN} = \frac{1}{2}\times (\texttt{GAINA} + \texttt{GAINB})
\label{eq::add-GAIN}
\end{equation}

\begin{equation}
\texttt{SATURATE} = \text{min}(\texttt{SATURATA}, \texttt{SATURATB})
\label{eq::add-SATURATE}
\end{equation}

%
%

 \end{appendix}

\bibliographystyle{pasa-mnras}
\bibliography{bibliography}

\newcommand{\noop}[1]{}
\begin{thebibliography}{}
\makeatletter
\relax
\def\mn@urlcharsother{\let\do\@makeother \do\$\do\&\do\#\do\^\do\_\do\%\do\~}
\definecolor{darkblue}{rgb}{0,0,0.597656}
\def\mndoi{\begingroup\mn@urlcharsother \@ifnextchar [ {\mndoi@} {\mndoi@[]}}
\def\mndoi@[#1]#2{\def\@tempa{#1}\ifx\@tempa\@empty \href
  {http://dx.doi.org/#2} {\textcolor{darkblue}{doi:#2}}\else \href
  {http://dx.doi.org/#2} {\textcolor{darkblue}{#1}}\fi \endgroup}
\def\mn@eprint#1#2{\mn@eprint@#1:#2::\@nil}
\def\mn@eprint@arXiv#1{\href {http://arxiv.org/abs/#1} {{\tt arXiv:#1}}}
\def\mn@eprint@dblp#1{\href {http://dblp.uni-trier.de/rec/bibtex/#1.xml}
  {dblp:#1}}
\def\mn@eprint@#1:#2:#3:#4\@nil{\def\@tempa {#1}\def\@tempb {#2}\def\@tempc
  {#3}\ifx \@tempc \@empty \let \@tempc \@tempb \let \@tempb \@tempa \fi \ifx
  \@tempb \@empty \def\@tempb {arXiv}\fi \@ifundefined
  {mn@eprint@\@tempb}{\@tempb:\@tempc}{\expandafter \expandafter \csname
  mn@eprint@\@tempb\endcsname \expandafter{\@tempc}}}

\bibitem[\protect\citeauthoryear{{Andreoni}, {Jacobs}, {Hegarty}, {Pritchard},
  {Cooke}  \& {Ryder}}{{Andreoni} et~al.}{in press}]{Andreoni-accepted}
{Andreoni} I.,  {Jacobs} C.,  {Hegarty} S.,  {Pritchard} T.,  {Cooke} J.,
  {Ryder} S.,  in press, \pasa

\bibitem[\protect\citeauthoryear{{Astropy Collaboration} et~al.,}{{Astropy
  Collaboration} et~al.}{2013}]{Astropy2013A&A...558A..33A}
{Astropy Collaboration} et~al., 2013, \mndoi [\aap]
  {10.1051/0004-6361/201322068}, \href
  {http://adsabs.harvard.edu/abs/2013A%26A...558A..33A} {558, A33}

\bibitem[\protect\citeauthoryear{Belmon}{Belmon}{1998}]{Belmon:Thesis:1998}
Belmon L.,  1998, PhD thesis, University of Paris XI Orsay, France

\bibitem[\protect\citeauthoryear{{Belmon}, {Benoit-Cattin}, {Baskurt}  \&
  {Bougeret}}{{Belmon} et~al.}{2002}]{Belmon2002A&A...386.1143B}
{Belmon} L.,  {Benoit-Cattin} H.,  {Baskurt} A.,   {Bougeret} J.-L.,  2002,
  \mndoi [\aap] {10.1051/0004-6361:20020225}, \href
  {http://adsabs.harvard.edu/abs/2002A%26A...386.1143B} {386, 1143}

\bibitem[\protect\citeauthoryear{Bradley et~al.,}{Bradley
  et~al.}{2016}]{larry_bradley_2016_155353}
Bradley L.,  et~al., 2016, astropy/photutils v0.2.2,
  \mndoi{10.5281/zenodo.155353}, \url {https://doi.org/10.5281/zenodo.155353}

\bibitem[\protect\citeauthoryear{{Brueckner} et~al.,}{{Brueckner}
  et~al.}{1995}]{Brueckner1995SoPh..162..357B}
{Brueckner} G.~E.,  et~al., 1995, \mndoi [\solphys] {10.1007/BF00733434}, \href
  {http://adsabs.harvard.edu/abs/1995SoPh..162..357B} {162, 357}

\bibitem[\protect\citeauthoryear{{Clark}}{{Clark}}{2008}]{Clark2008}
{Clark} A.,  2008, \mndoi [IEEE Signal Processing Magazine]
  {10.1109/MSP.2008.918681}, \href
  {http://adsabs.harvard.edu/abs/2008ISPM...25..146C} {25, 146}

\bibitem[\protect\citeauthoryear{Cohen, Daubechies  \& Feauveau}{Cohen
  et~al.}{1992}]{cohen1992biorthogonal}
Cohen A.,  Daubechies I.,   Feauveau J.-C.,  1992, Communications on pure and
  applied mathematics, 45, 485

\bibitem[\protect\citeauthoryear{{Diehl} \& {Dark Energy Survey
  Collaboration}}{{Diehl} \& {Dark Energy Survey
  Collaboration}}{2012}]{Diehl2012PhPro..37.1332D}
{Diehl} T.,  {Dark Energy Survey Collaboration} 2012, \mndoi [Physics Procedia]
  {10.1016/j.phpro.2012.02.472}, \href
  {http://adsabs.harvard.edu/abs/2012PhPro..37.1332D} {37, 1332}

\bibitem[\protect\citeauthoryear{{Flaugher} et~al.,}{{Flaugher}
  et~al.}{2012}]{Flaugher2012SPIE.8446E..11F}
{Flaugher} B.~L.,  et~al., 2012, in Ground-based and Airborne Instrumentation
  for Astronomy IV. p. 844611, \mndoi{10.1117/12.926216}

\bibitem[\protect\citeauthoryear{{Flaugher} et~al.,}{{Flaugher}
  et~al.}{2015}]{Flaugher2015AJ....150..150F}
{Flaugher} B.,  et~al., 2015, \mndoi [\aj] {10.1088/0004-6256/150/5/150}, \href
  {http://adsabs.harvard.edu/abs/2015AJ....150..150F} {150, 150}

\bibitem[\protect\citeauthoryear{Folk, Heber, Koziol, Pourmal  \&
  Robinson}{Folk et~al.}{2011}]{Folk:2011:OHT:1966895.1966900}
Folk M.,  Heber G.,  Koziol Q.,  Pourmal E.,   Robinson D.,  2011, in
  Proceedings of the EDBT/ICDT 2011 Workshop on Array Databases. AD '11.
ACM, New York, NY, USA, pp 36--47, \mndoi{10.1145/1966895.1966900}, \url
  {http://doi.acm.org/10.1145/1966895.1966900}

\bibitem[\protect\citeauthoryear{{ISO/IEC 15444-1:2000}}{{ISO/IEC
  15444-1:2000}}{2000}]{JPEG2000-part1}
{ISO/IEC 15444-1:2000} 2000, Technical report, {Information technology -- JPEG
  2000 image coding system -- Part 1: Core coding system}.
{ISO/IEC}

\bibitem[\protect\citeauthoryear{Kitaeff, Cannon, Wicenec  \& Taubman}{Kitaeff
  et~al.}{2015}]{Kitaeff2015229}
Kitaeff V.,  Cannon A.,  Wicenec A.,   Taubman D.,  2015, \mndoi [Astronomy and
  Computing] {http://dx.doi.org/10.1016/j.ascom.2014.06.002}, 12, 229

\bibitem[\protect\citeauthoryear{{Klebesadel}, {Strong}  \&
  {Olson}}{{Klebesadel} et~al.}{1973}]{Klebesadel1973ApJ...182L..85K}
{Klebesadel} R.~W.,  {Strong} I.~B.,   {Olson} R.~A.,  1973, \mndoi [\apjl]
  {10.1086/181225}, \href {http://adsabs.harvard.edu/abs/1973ApJ...182L..85K}
  {182, L85}

\bibitem[\protect\citeauthoryear{Labrum, McLean  \& Wild}{Labrum
  et~al.}{1975}]{labrum1975radioheliography}
Labrum N.,  McLean D.,   Wild J.,  1975, in Methods in Computational Physics.
  Volume 14-Radio astronomy. pp 1--53

\bibitem[\protect\citeauthoryear{Li}{Li}{2003}]{li-2003}
Li J.,  2003, Modern Signal Processing, 46, 185

\bibitem[\protect\citeauthoryear{{Li} \& {Paczy{\'n}ski}}{{Li} \&
  {Paczy{\'n}ski}}{1998}]{Li1998ApJ...507L..59L}
{Li} L.-X.,  {Paczy{\'n}ski} B.,  1998, \mndoi [\apjl] {10.1086/311680}, \href
  {http://adsabs.harvard.edu/abs/1998ApJ...507L..59L} {507, L59}

\bibitem[\protect\citeauthoryear{{Lorimer}, {Bailes}, {McLaughlin}, {Narkevic}
  \& {Crawford}}{{Lorimer} et~al.}{2007}]{Lorimer2007Sci...318..777L}
{Lorimer} D.~R.,  {Bailes} M.,  {McLaughlin} M.~A.,  {Narkevic} D.~J.,
  {Crawford} F.,  2007, \mndoi [Science] {10.1126/science.1147532}, \href
  {http://adsabs.harvard.edu/abs/2007Sci...318..777L} {318, 777}

\bibitem[\protect\citeauthoryear{{Meade} et~al.,}{{Meade}
  et~al.}{2017}]{Meade2017arXiv170401281M}
{Meade} B.,  et~al., 2017, \mndoi [\pasa] {10.1017/pasa.2017.15}, \href
  {http://adsabs.harvard.edu/abs/2017PASA...34...23M} {34, e023}

\bibitem[\protect\citeauthoryear{{Miller}, {Fabian}  \& {Miller}}{{Miller}
  et~al.}{2004}]{Miller2004ApJ...614L.117M}
{Miller} J.~M.,  {Fabian} A.~C.,   {Miller} M.~C.,  2004, \mndoi [\apjl]
  {10.1086/425316}, \href {http://adsabs.harvard.edu/abs/2004ApJ...614L.117M}
  {614, L117}

\bibitem[\protect\citeauthoryear{{Morii}, {Ikeda}, {Sako}  \& {Ohsawa}}{{Morii}
  et~al.}{2017}]{Morii2017ApJ...835....1M}
{Morii} M.,  {Ikeda} S.,  {Sako} S.,   {Ohsawa} R.,  2017, \mndoi [\apj]
  {10.3847/1538-4357/835/1/1}, \href
  {http://adsabs.harvard.edu/abs/2017ApJ...835....1M} {835, 1}

\bibitem[\protect\citeauthoryear{{Pence}, {White}, {Greenfield}  \&
  {Tody}}{{Pence} et~al.}{2000}]{Pence2000}
{Pence} W.,  {White} R.~L.,  {Greenfield} P.,   {Tody} D.,  2000, in {Manset}
  N.,  {Veillet} C.,   {Crabtree} D.,  eds,  Astronomical Society of the
  Pacific Conference Series Vol. 216, Astronomical Data Analysis Software and
  Systems IX. p.~551

\bibitem[\protect\citeauthoryear{{Pence}, {White}  \& {Seaman}}{{Pence}
  et~al.}{2010}]{Pence2010PASP..122.1065P}
{Pence} W.~D.,  {White} R.~L.,   {Seaman} R.,  2010, \mndoi [Publications of
  the Astronomical Society of the Pacific] {10.1086/656249}, \href
  {http://adsabs.harvard.edu/abs/2010PASP..122.1065P} {122, 1065}

\bibitem[\protect\citeauthoryear{{Pence}, {Seaman}  \& {White}}{{Pence}
  et~al.}{2011}]{Pence2011}
{Pence} W.,  {Seaman} R.,   {White} R.~L.,  2011, in {Evans} I.~N.,
  {Accomazzi} A.,  {Mink} D.~J.,   {Rots} A.~H.,  eds,  Astronomical Society of
  the Pacific Conference Series Vol. 442, Astronomical Data Analysis Software
  and Systems XX. p.~493

\bibitem[\protect\citeauthoryear{{Peters} \& {Kitaeff}}{{Peters} \&
  {Kitaeff}}{2014}]{Peters-2014}
{Peters} S.~M.,  {Kitaeff} V.~V.,  2014, \mndoi [Astronomy and Computing]
  {10.1016/j.ascom.2014.06.003}, \href
  {http://adsabs.harvard.edu/abs/2014A\%26C.....6...41P} {6, 41}

\bibitem[\protect\citeauthoryear{Rabbani \& Joshi}{Rabbani \&
  Joshi}{2002}]{Rabbani20023}
Rabbani M.,  Joshi R.,  2002, \mndoi [Signal Processing: Image Communication]
  {https://doi.org/10.1016/S0923-5965(01)00024-8}, 17, 3

\bibitem[\protect\citeauthoryear{{Rest} et~al.,}{{Rest}
  et~al.}{2005}]{Rest2005ApJ...634.1103R}
{Rest} A.,  et~al., 2005, \mndoi [\apj] {10.1086/497060}, \href
  {http://adsabs.harvard.edu/abs/2005ApJ...634.1103R} {634, 1103}

\bibitem[\protect\citeauthoryear{Rissanen \& Langdon}{Rissanen \&
  Langdon}{1979}]{Rissanen5390830}
Rissanen J.,  Langdon G.~G.,  1979, \mndoi [IBM Journal of Research and
  Development] {10.1147/rd.232.0149}, 23, 149

\bibitem[\protect\citeauthoryear{Shannon}{Shannon}{1948}]{shannon}
Shannon C.,  1948, Bell System Technical Journal, Vol. 27, 379

\bibitem[\protect\citeauthoryear{{Smithsonian Astrophysical
  Observatory}}{{Smithsonian Astrophysical
  Observatory}}{2000}]{ds9-2000ascl.soft03002S}
{Smithsonian Astrophysical Observatory} 2000, {SAOImage DS9: A utility for
  displaying astronomical images in the X11 window environment}, Astrophysics
  Source Code Library (\mn@eprint {ascl} {0003.002})

\bibitem[\protect\citeauthoryear{{Tomasi}}{{Tomasi}}{2016}]{Tomasi2016A&C....16...88T}
{Tomasi} M.,  2016, \mndoi [Astronomy and Computing]
  {10.1016/j.ascom.2016.04.004}, \href
  {http://adsabs.harvard.edu/abs/2016A%26C....16...88T} {16, 88}

\bibitem[\protect\citeauthoryear{{Valdes}, {Gruendl}  \& {DES
  Project}}{{Valdes} et~al.}{2014}]{Valdes2014ASPC..485..379V}
{Valdes} F.,  {Gruendl} R.,   {DES Project} 2014, in {Manset} N.,  {Forshay}
  P.,  eds,  Astronomical Society of the Pacific Conference Series Vol. 485,
  Astronomical Data Analysis Software and Systems XXIII. p.~379

\bibitem[\protect\citeauthoryear{Vohl}{Vohl}{2013}]{Vohl:Thesis:2013}
Vohl D.,  2013, Master's thesis, Universit\'e Laval, Canada

\bibitem[\protect\citeauthoryear{Vohl, Fluke  \& Vernardos}{Vohl
  et~al.}{2015}]{Vohl2015200}
Vohl D.,  Fluke C.,   Vernardos G.,  2015, \mndoi [Astronomy and Computing]
  {http://dx.doi.org/10.1016/j.ascom.2015.05.003}, 12, 200

\bibitem[\protect\citeauthoryear{{Wells}, {Greisen}  \& {Harten}}{{Wells}
  et~al.}{1981}]{Wells1981A&AS...44..363W}
{Wells} D.~C.,  {Greisen} E.~W.,   {Harten} R.~H.,  1981, Astronomy and
  Astrophysics Supplement Series, \href
  {http://adsabs.harvard.edu/abs/1981A%26AS...44..363W} {44, 363}

\bibitem[\protect\citeauthoryear{{White} \& {Percival}}{{White} \&
  {Percival}}{1994}]{white-1994}
{White} R.~L.,  {Percival} J.~W.,  1994, in {Stepp} L.~M.,  ed.,  Society of
  Photo-Optical Instrumentation Engineers (SPIE) Conference Series Vol. 2199,
  Advanced Technology Optical Telescopes V. pp 703--713

\bibitem[\protect\citeauthoryear{Wu \& Chin}{Wu \& Chin}{2014}]{Wu2014BDR}
Wu Z.,  Chin O.~B.,  2014, \mndoi [Big Data Research]
  {http://dx.doi.org/10.1016/j.bdr.2014.08.002}, 1, 1

\makeatother
\end{thebibliography}

\end{document}